\documentstyle[12pt]{article}

\def\be{\begin{equation}}
\def\ee{\end{equation}}
\def\a{\alpha}
\def\d{\delta}
\def\D{\Delta}
\def\g{\gamma}
\def\G{\Gamma}
\def\kt{{\tilde k}}
\def\l{\lambda}
\def\o{\omega}
\def\da{\dagger}
\def\bg{b^{\dagger}}
\def\cg{c^{\dagger}}
\def\s{\sigma}
\def\su{\underline {\sigma}}
\def\r{\rightarrow}
\def\e{\epsilon}
\def\ur{\uparrow}
\def\dr{\downarrow}
\def\f{\varphi}

\begin{document}
\vspace{1.5in}
\begin{center}

{\Large \bf Interplay of pseudo- and superconducting
gaps: observable manifestations in cuprates. } \\

\vspace{0.3in}
{\large \bf A.A.~Ovchinnikov and M.Ya.Ovchinnikova. }   \\
\vspace{0.2in}
{\it Institute of Chemical Physics of RAS, Moscow\\
MP Institute of Physics of Complex systems, Dresden. } \\
\vspace{0.3in}
\end{center}

\begin{abstract}

Angular dependence of gap, seen in photoemission, its evolution with
doping and temperature are interpreted on base t-t'-U Hubbard model
in which a pseudogap is a working function for electrons removing
from dielectric segments of zone boundary. Tunnel spectra in one-particle
approximation display only superconducting gap and too large asymmetry
in regions aside from optimal doping. Spectral functions confirm sharp
well defined Fermi boundary at $k_x=k_y$ direction and smoothed destructed
one at the $\G(0,0)-M(\pi,0)-Y(\pi ,\pi )$ path. Both tunnel and ARPES data
seem require spreading a model to bilayer ones.

\end{abstract}

\vspace{0.12in}

\vspace{0.12in}
{\bf 1. Introduction.}
\vspace{0.12in}

Up to now a set of decisive experimental results have been obtained
concerning electronic structure of cuprates (see reviews \cite{1,2,3}).
They include evidence of d-wave superconductivity,
discovery of anisotropic normal state pseudogap  and
"small" Fermi surface (FS) in underdoped (UD) materials accompanied
possibly by destruction of FS \cite{4,5}. Direct measurement of
superconducting (SC) gap are provided by tunneling spectroscopy
\cite{6}-\cite{8a}. Most recent
detailed photoemission (ARPES) studies  reveal
exotic temperature dependence of gap \cite{9,10} and the doping
dependence of its anisotropy in superconducting state \cite{11}.
It was revealed the apparent discrepancy between
this data and characteristics obtained from penetration depth data.
It was shown that angular dependence of superconducting
gap $\D (\f)$ extracted from ARPES data cannot be adjusted by simple
form $\sim \cos (2\f )$ and the shape of gap function
depends on doping. In particular, the ratio $\xi =v_\D /\D $ of $v_\D
={1\over 2} d\D (\f ) /d\f$ at  $\f=\pi/4$ to maximum value of gap
$\D$ decreases
essentially with doping decrease in UD materials. Finally standard
shape of the FS has been recently revised. It seems to be large partially
electron-like FS around $\G =(0,0)$ instead the arc around $Y(\pi , \pi )$
\cite{12}.

   The aim of our paper is to study which of new experiments can or cannot be
explained in simple quasiparticle interpretation of pseudogap origin.

Interpretation of phase diagram and pseudogap
have recently  \cite{13}-\cite{17}
been proposed on base of different approaches for
description of the correlation mechanism of SC pairing in cuprates.
One of them is the spin-polaron approach in t-t'-J model \cite{13}.
Another is variational approach incorporating the correlations of the
valence bond (VB) type in t-t'-U Hubbard model \cite{14,15}. In each case
the Hubbard splitting of whole band arises due to long-range antiferromagnet
(AF) order and fine details of upper edge of lower Hubbard band
(LHB) determines the low energy phenomena. The hole attraction in d-channel
is shown to be induced by spin-polaron interactions in t-t'-J model or
equivalent correlated hopping interactions due to VB formation in t-t'-U
model. According theories \cite{13}-\cite{17} the next nearest hopping t'
plays important role in formation of pseudogap and a sign of t'/t determines
its anisotropy. For systems with $t'/t>0$ a crossing of van Hove singularity
(VHS) by the chemical potential $\mu$ corresponds to optimal doping and is
accompanied by change in the Fermi boundary topology. The hole
pockets around $k=(\pi /2, \pi /2)$ with "small" Fermi surface are replaced
by large Fermi surface in course of transition from UD to over-doped (OD)
systems. At $\d <\d_{opt}$ the VHS occur below $\mu$ and some segments
of magnetic Brillouin zone (MBZ) boundary
near $k_M=(\pi,0)$ become dielectric
segments. According \cite{15}  the work function $\mu-E_k>0$ for
electron removing from this regions of $k$ is just a normal state
pseudogap observed in ARPES. In difference from standard VHS scenario
\cite{3} with united band here we deal with VHS of lower Hubbard
subband.

The method and outline of present study are following.
We use the variational method to construct explicitly
the correlated state and to test if it could display the observable
properties. Starting from AF state with the VB correlations we
explain the superconducting (SC) and  pseudo- gap anisotropy and
its temperature and doping dependencies (Sec.2). Then we calculate the
tunnel spectra of system and discuss its distinction from the observed ones
(Sec.3). Manifestation of FS in one-particle spectral
functions $n_k$ and $A(k,\o\r 0)$ and the FS shape are discussed in Sec.4.

\vspace{0.12in}
{\bf 2. The doping and temperature dependence of gap anisotropy.}
\vspace{0.12in}

Calculations have been done for t-t'-U-V Hubbard model by method proposed in
\cite{15}. Here $t,~t'$ are the nearest and next nearest hoppings; U
and V are the on-site and nearest sites Coulomb-like interactions.
The variational correlated state $\Psi=W(\a)\Phi$ with the VB-type
correlations (the band analogue of the Anderson's RVB
states) is taken as a unitary transformed uncorrelated state $\Phi$.
Unitary operator $W(\a)=\exp\{\a \sum_{<nm>\s} (\cg_{n\s}c_{m\s}-h.c.)
(n_{n\su}-n_{m\su})\}$ , $\su =-\s$, optimizes the covalency of $<nm>$
bonds. A problem with Hamiltonian $H$ in basis of correlated states
$\{\Psi\}$ is reduced to a problem with effective
Hamiltonian ${\tilde H} =W^\da HW \approx H +\a [H, Z] +{\a^2\over 2}
[[H, Z], Z]$ in basis of uncorrelated states ${\Phi}$. Thus the effective
problem ${\tilde H}$  is solved in the mean field (MF) approximation
with subsequent minimization of energy over variational parameter
$\a$ ($\a <0.22$ at $U/t\le 9$). Most general uncorrelated state
${\Phi}$ of BCS-type and with double magnetic unit cell allows a testing
the possible AF and SC pairing in system. The terms of correlated
hopping like $\sim \a U \cg_{n\s}c_{m\s}n_{n\su}n_{m\su}$ in ${\tilde H}$
caused by VB formation are responsible for SC order of d-symmetry
compatible with AF order. Band energies $E_{k\l}$ (eigen values of linearized
${\tilde H}_L$) corresponds to self consistent $\Phi$ and optimal $\a$.

The studied homogeneous solutions display the 2D AF spin order in a
wide doping range $\d=|1-n|\le\d_c\sim 0.3$ which is larger
than range of the bulk AF order ($\d_c\sim 0.05$).
This implies that finite radius of AF correlations is much
larger than lattice constant $a$ and we can use the solutions and a
picture of split band as a background in discussing a low energy
phenomena. In normal state a LHB has the following
structure \cite{15}:
\be
E_k = [\e_0+4t'c_xc_y+...]-\sqrt{[Ud_0+...]^2+[2t(c_x+c_y)+...]^2}
\label{1}
\ee
Here only main contributions from lowest harmonics with $c_{x(y)}=
\cos{k_{x(y)}}$ are explicitly written and $d_0=<(-1)^nS_{zn}>$ is the
staggered spin. In d-wave superconducting state the self consistent
quasipartilce energies are very close to form
\be
E_{k\l}= \pm \sqrt{(E_{k}-\mu)^2+W_k^2};~~~
W_k\sim F_k[k_{11}(c_x-c_y)w_1+...]
\label{2}
\ee
Here $W_k$ is determined by SC order parameters
$w_l=(l_x^2-l_y^2)<\cg_{n\ur } \cg_{n+l,\ur }>$,
$|l|=1,~\sqrt{5},~3$ and $F_k$  is determined by structure
of one-particle states of LHB. First harmonic gives main contribution
to $W_k$ and its angular dependence is close to
$W_k\sim (c_x-c_y)$.

The doping dependence $T_c(\d)$ for Hubbard model with $U/t=8$,
$t'/t=0.05$ and
$V/t=0.1$ is presented in Fig.1a. (Note, that at V=0 the value of
$T_c^{max}$ increases by factor 1.64 \cite{15}.) Curve 1 in Fig.1b presents
the value $W_M(\d)=W(k_M)$ of gap function $W_k$ at $k=k_M=(\pi,0)$
for the d-wave
superconducting pairing calculated at temperature $T/t=0.002$. But according
(2) the quasiparticle energy and the corresponding shift $\D=\D\o$ of the
edge of photoemission distribution function at $k=(\pi,0)$ is determined not
only by $W_k$, but also by the LHB energy $|E_k-\mu|$ relative to chemical
potential. Unlike original unsplit band, the LHB energy $E_k$ is periodic
in magnetic Brillouin zone (MBZ) and goes through a maximum on MBZ boundary
at any path connecting the points $\G (0 ,0)$ and $Y(\pi ,\pi )$.
This determines the different doping dependence of ratio
$\xi =v_\D/\D_M$ in UD and OD presented in Fig.1c.
Here $v_{\D}={1\over 2}d\D/d\f$ at $\f =\pi /4$ and $\D_M$ is the
$\D$ at $k=(\pi ,0)$.

   In OD regime each path $\G-Y$ crosses a "large"
Fermi-boundary: $E_k-\mu=0$ at $k=k_F$ and $E_k-\mu>0$ at all values $k$
at MBZ boundary. In this case a gap $\D(\f )$ in excitation spectrum for $k$
changing along any path $\G -Y$ is determined only by d-wave SC gap
$W_k$ at $k_F$. Therefore the gap anisotropy $\D(\f )$ coincides with
$\D(\f )=W_k(\f )\sim z(\f )=0.5(\cos{k_x}-\cos{k_y})$ at $k_F$.
Corresponding ratio
$\xi=v_\D/\D_M$ measured at $k=k_M$ in \cite{11} is equal to
$\xi={1\over 2}(dW_k/d\f )/W_M$ . It is close to
constant value $\sim 0.7$ expected for pure d-wave SC order  without
pseudogap.

In UD systems with $t'/t>0$ a "small" Fermi boundary around
the hole pockets at $k=(\pi /2,\pi /2)$ is formed whereas the segments of
MBZ boundary near
$M(\pi ,0)$ become dielectric segments on which $E_k-\mu<0$.  A
shift $\D(k_M)$ of photoelectron distribution function in ARPES
at $k_M=(\pi ,0)$ is equal to $\D(k_M)=\sqrt{\D_N^2+W^2_{M}}$.
Here $W_M$ is $W_k$ at $k_M$ and the value  $\D_N=|E_{k}-\mu|$ at $k_M$
is the normal state pseudogap equal to work function for the electron
removing from $k=k_M$ in normal state. Value $\D_N$ increases with doping
decrease. For the same reason the anisotropy of total shift $\D(\f )=
\sqrt{\D_N^2(\f )+W_k^2(\f)}$ of photoemission edge at the dielectric
segments of MBZ boundary differs significantly from anisotropy
of $W_k(\f )$ only. In UD region ($\D_N\neq 0$) one obtains
$\xi\sim 0.5[W'(\pi /4)]/\sqrt{\D_N^2+W_M^2}$, where
$W' (\f)=dW_k/d\f$ at $\f=\pi /4$. Value $\xi$
sharply decreases ($W_k\r 0$) with doping decrease
in full accordance with experimental
behavior of $\xi (\d )$ \cite{11}. In contrast to $\xi(\d)$,
a quantity $\xi _W(\d )=v_\D /W_M$ with pure superconducting gap
function in denominator remains almost constant (curve 2 in Fig.1c).

Fig.1d presents the results for anisotropic shift $\D(\f )$ of
photoemission edge for strongly UD system ($\d=0.14$,
$T_c/T_c^{max}=0.55$).  Dependencies $\D (\f )$ are taken at $k$ moving from
$k_x=k_y$ to $M(\pi ,0)$ along generalized Fermi boundary consisting of
nonshadow part of small Fermi boundary and dielectric segments of MBZ (see
\cite{15}). Several features of calculated anisotropy $\D (\f )$ have a
correspondence with experimental anisotropy features. 1) Deflection of
$\D (\f )$ from linear dependence $\D (\f )\sim z(\f )$
have really been observed in UD Bi-Sr-Ca-Cu-O (BSCCO) \cite{11} and the
observed sign of next harmonics in it is the same as in
calculated curves. 2) The region of gap $\D_k$ near its nodes occurs
more sensitive to
temperature than at $k\sim (\pi,0)$. A shift $\D (\f )$ at $\f\sim \pi /4$
disappears with $T\r T_c$ whereas the pseudogap of normal state in region
$k\sim (\pi,0)$ retains nonzero value $\D_M=\D_N=\mu-E(\pi,0)>0$ at $T>T_c$.
This feature also corresponds to experimentally observed different temperature
sensitivity of $\D (\f )$ at different $\f$ \cite{9,10}. Finite
$k$-resolution in ARPES may be a reason of more smooth observed dependence
$\D (\f)$ in compare with calculated one. Besides, the latter is
obtained in neglecting the energy dispersion over $k_z$. 3) The
calculated ratio $\eta =2\D_M/kT_c$ have very large values in underdoped
case. Thus for $\d=0.16 $ ($T_c=0.9 T_c^{max}$) it is $\eta =9.5$. At the
same time for optimal and over- doping the whole gap is determined only by
superconducting part $\D =W(k_F)$ and the ratio is about $\eta \approx 5$.
This value is close to $\eta =4.5$ obtained from empirical BCS
models with d-wave pairing. Decrease of $2\D/kT_c$ with doping
increase is consistent with experiment \cite{1}.

Independent estimations of gap anisotropy near the nodes have been obtained
in \cite{11} from behavior of superfluid density $\rho_s$ at low temperature
on base of relation  between $v_{\D}$ and penetration length at $T\r 0$.
This relation $d\rho_s (T)/dT\sim d(\l^{-2})/dT\sim 1/v_{\D}$ gives the doping
dependence of $v_{\D}$ different from the mutually consistent dependencies
$v_{\D}(\d )$ obtained from ARPES data and our calculations. Origin of
distinction is unclear and question arises whether the temperature
dependence of $1/\l^2$ at $T\r 0$ may be controlled not only by
low energy Fermi excitations near the nodes  of SC gap, but also by the low
frequency spin excitations which adiabatically change the overall spectrum of
Fermi system. Important role of spin wave fluctuations has recently been
confirmed in \cite{19}.

\vspace{0.12in}
{\bf 3. Tunneling spectra for AF+VB correlated states}
\vspace{0.12in}

Tunneling spectra can provides a sensitive test for the obtained simple
homogeneous solutions  of $t-t'-U$ Hubbard model which take into account
the AF and VB correlations. Recent reliable experimental results in tunnel
spectroscopy \cite{5}-\cite{8a} give the base for such testing.

Consider so-called SIN (superconductor-insulator-normal metal) tunneling
typical for STM measurements. Total tunnel current
$I(V)=e(W_I-W_{II})$ is determined by probabilities $W_{I(II)}$ of electron
transfer from superconductor to normal metal and vice versa. These
probabilities are found in first order in tunnel interaction
$V_{tun}\sim \sum_{\s,n} G(n-n_0)(\cg _{n\s}d_{\e\s}+h.c.)$.
Here $d _{\e\s}$ are the Fermi operators of normal metal (a tip of STM).
A "tip formfactor" $G(n-n_0)$ characterizes the extension of surface-tip
interaction centered at some site $n_0$. As usually the constant
density of state (DOS) for normal
metal (tip of STM) and the energy independent formfactor are assumed.

For state with AF+VB correlations and d-wave SC order the quasiparticle
operators $\chi^\da _{k\l}=\bg_{ki}U_{i\l}$, $~i,\l=1,\ldots ,4$,
are eigen operators of linearized effective
Hamiltonian ${\tilde H}_L$ \cite{15}. Here a basis set of Fermi operators is
$\bg _{ki}=\{\cg _{k\ur},\cg_{\kt \ur},c_{-k\dr},c_{-\kt \dr}\}_i$,
$\kt =k-(\pi,\pi )$, and
$k\in F$ varies inside MBZ $|k_x\pm k_y|\leq \pi$. With taking into account
the structure and spectrum $E_{k\l}$ of quasiparticles the differential
tunnel current takes the form
\be
{{dI(V)} \over {dV}}={-1 \over N}{\sum_k}^F \Bigl\{
[R U_{1\l}^2+{\tilde{R}}U_{2\l}^2] f' (E_{k\l}-eV)
+[R U_{3\l}^2+{\tilde{R}}U_{4\l}^2] f' (E_{k\l}+eV)
\Bigr\}
\label{3}
\ee
Here $R=G^2(k)$, ${\tilde{R}}=G^2(\kt )$ and
$f '(E)=df/dE$ is derivative of Fermi function.
Deducing of Eq.(5) includes the averaging over the position of "tip
center" $n_0$ among two magnetic sublattices of AF state. Sign of $V$
is chosen so that negative V corresponds to occupied levels in the
sample spectrum. The matrix elements $U_{i\l}$ are the analogues of the
coherence factors in simple BCS theory. As in \cite{20} they are retained
in explicit form in Eq.(5) since the normal state DOS of
superconductor sharply depends on energy near VHS. Note, that here we deals
with VHS of LHB instead of VHS of whole unsplit band as in \cite{20}.

Figures 2,3 present the tunnel spectra (3) for nearly optimal doping
and for UD and OD systems at temperatures lower and
higher $T_c$ for one of studied variants of "tip formfactor".
In complete accordance with expectation at
temperature lower than $T_c$ the SC gap (not pseudogap) displays in
differential tunnel current. The distance between peaks  asides main
dip at V=0 corresponds to gap $2\D_{SC}(\d)$ whose values and
doping dependence at small $T$ are close to dependence $2W_M (\d )$ in
Fig.1b and is characterized by maximum at optimal doping.

In OD region the calculated gap behavior  $2\D_{SC}(\d)$ is in
accordance with gap behavior measured by STM in
BSCCO \cite{8a}. In both cases the gap
decreases with  doping at $\d >\d_{opt}$. In UD samples the observed
tunnel gap seems to include the pseudogap
which does not vanish at $T>T_c$ \cite{8,1}. This result is not
reproduced by calculations. Besides, at all doping the observed spectra
retain almost symmetrical form close to calculated spectrum for optimal
doping (Fig.2). However a tunnel current $dI/dV$ becomes highly
asymmetric in UD and OD systems. At low temperatures sharp peak in function
$dI/dV$ at $V<0$ or $V>0)$ for UD and OD doping correspondingly reflects the
VHS in the normal state DOS. The asymmetry of observed spectra remains almost
invariable at slight variations of doping arround optimal value \cite{6,7}.
The shape of DOS determines also asymmetric
nonlinear background of tunnel current at large $V$ (see inset in Fig.2).
However the one-particle approximation (3) used is unsuitable at large $V$.
It cannot describe
secondary wide peaks in observed tunnel spectra at large V. These peaks
(as well as similar patterns in ARPES data) are connected by many authors
with spin-wave excitations \cite{21}. Possible origin of VHS broadening or splitting
removing too sharp asymmetry may consist in interlayer coupling of $CuO_2$
planes. We neglect this coupling. The split of VHS in bilayer system and
position of $\mu$ between two VHS might manifest as two almost symmetric
peaks in tunnel spectra which retains as pseudogap even at $T>T_c$.

Additional foundation for hypothesis on the bilayer-split have been
obtained recently from dependencies of spectral functions $A(k,\o)$ on
photon energy in ARPES data \cite{12}. The observed spectral features
are discussed in terms of united band without huge gap between Hubbard
subband which are implied in our calculations. One must also keep in mind
possible inequivalence of $CuO_2$ planes closest to surface and those
in bulk caused by uncompensated fields of the charged BiO plane at
cleavage in STM experiment.

\vspace{0.12in}
{\bf 4. Manifestation of Fermi surface and influence of spin fluctuations.}
\vspace{0.12in}

One of open question in correlated variational approach [15] was how to
concert the ARPES data with predicted shape of nonshadow FS for OD
models. For such models "large" FS around point $\G (0,0)$ with electron-
like segments near $(\pi,0)$ was obtained. This was in contradiction with
widely recognized shape of FS as an arc around $Y(\pi,\pi)$ \cite{2}.
But recent new reexamination of FS shape for BSCCO shows
the possibility of the electron-like FS crossing on line $\G-M(\pi,0)$.
Besides the FS shape, it is important to discuss how various sections of
FS should manifest in spectral functions.

The second question is about a role of spin fluctuations. Important
influence of low frequency collective spin-wave  on Fermi spectrum
of correlated system is wide spread known (see \cite{19} among last
publication). Since the spin-wave states are the states with a spiral spin
structure, the comparative study of the normal state DOS and the band
structure features for AF and spiral states of $t-t'-U-V$ Hubbard model
have been carried out here. This study has only qualitative character.
For this reason and for sake of simplicity we consider and compare the
mean-field (MF) solutions with AF and spiral spin structures but
without correlations of the VB type.

Standard MF spiral state $\Psi_Q$ with a spirality vector $Q$  is the
MF state with one-electron averages $<\cg_{n\ur}c_{n\dr}>=b_0\exp[iQn]$.
At $Q\r Q_{AF}=(\pi,\pi)$ it comes to AF state.
Neglecting of VB correlations significantly changes the energy of
system and shifts to larger value a critical doping at which the MF AF (or
spiral) solutions transform to paramagnet ones \cite{15}. For the
same reason minimization of MF energy over $Q$ at given $\d$ leads to
overestimated difference between optimal vector  $Q(\d )$ and
$Q_{AF}=(\pi ,\pi )$. In this connection we study and compare the DOS
of spiral states at arbitrary vector Q, which does not minimize
the MF energy at each doping.

For MF states without VB correlations and SC pairing we define under-
and over- doped states as that for which the VHS occurs
lower or higher than the chemical potential $\mu$.
Then at "optimal" doping the VHS is just placed at chemical potential.
For calculated model with parameters $U/t=8$, $t'/t=0.05$, $V/t=0.1$,
such definition gives a value $\d_{opt}\sim 0.25$ for the "optimal" doping
in case of AF state. It is larger than value $\d_{opt}=0.19$ for the
same system in the AF+VB state which takes into account the VB correlations.

Fig.4 presents the DOS of LHB for MF AF and spiral states  with different
directions of $Q$  and $|Q-Q_{AF}|=0.3$. The single VHS for AF state
transforms to  split VHS's for spiral states. Formation of spiral spin
structure removes degeneracy of VHS at $k=(\pm \pi , \pm \pi )$ in the
same manner as the lattice distortion removes it in usual Jahn-Teller
effect \cite{3}. Spiral states provide examples of electronic analogues
of Jahn-Teller effect.

It is instructive to compare the influence of spin structure on occupation
number $n_k={1\over 2}\sum_\s <\cg_{k\s}c_{k\s}>$ in momenta space and
on the one-particle spectral function $A(k,\o )$ at low frequency $\o\r 0$.
\be
A(k,\o=0)={1\over {ZN}}\sum_{\s}\sum_{i,f}|<f|c_{k\s}|i>|^2
\exp(-E_i/kT)
{\g\over {\pi [(E_i-E_f-\o)^2+\g^2]}}\Biggr|_{\o =0}
\label{4}
\ee
Here $E_i,~E_f$ are energies of initial and final states, $Z$ is partitition
function of system, and standard $\d$-function $\d (E_i-E_f-\o )$
is replaced by Lorentz function with width $\g$.

Fig.5 presents functions $n_k$ and  $A(k,\o=0 )$ calculated
for k changing along the counter $Y(\pi ,\pi )-\G(0,0)-M(\pi, 0)-Y$
for under-, optimally and over- doped MF states of $t-t'-U-V$ model at
$\g =0.003t$.

Several features, partly well known, of obtained spectral functions must
be stressed.

Even in AF state, for which the band energy are periodic in MBZ,
functions $n_k$, $A(k,\o )$  do not possess such periodicity. Neglecting
fine details, one can see that occupied ($n_k>0.5$) or empty ($n_k<0.5$)
momenta states refer to regions around $\G$ or $Y$ correspondingly, just in
same manner as it takes place for noninteracting system. This property is
well known and retains for spiral states also.

At $\G -Y$ direction the most sharp fall of $n_k$ is seen at
$k_x=k_y<\pi /2$ and corresponds to crossing of nonshadow Fermi boundary
with line $\G-Y$. Position of fall corresponds to the standard observed Fermi
boundary in form of arc with center at $Y(\pi,\pi)$. Involving the spiral
(instead of AF) spin order does not change the position of Fermi boundary
at this direction. Corresponding peak in $A(k,\o=0 )$ is most prominent
at any doping and does not influenced by spin fluctuations, i.e. its position
does not depend on the spirality vector $Q$.
This is in accordance with the observed
well defined Fermi surface at $\G -Y$ direction for any doping
\cite{4}. The second peak of  $A(k,\o=0 )$ on the same direction and
corresponding step in $n_k$ refers to shadow Fermi boundary. The intensity
of peak is much less and its position significantly depends on $Q$.
So this feature is expected to be smoothed by spin fluctuations.

At any doping in region near $M(\pi,0)$ the occupation number $n_k$
changes smoothly with k running along $\G -M-Y$ and low-frequency Fermi
excitation peaks in $A(k,\o=0 )$ are sufficiently intensive in whole
region near $M$ at any doping, especially at "optimal" doping (Fig.5).
This is the same in both cases: at UD when the region near $M$
become dielectric one ($E(\pi ,0)-\mu<0$) and at OD when the region
near M arranges between main and shadow large FS's ($E(\pi ,0)-\mu>0$).
Similar smoothing or destruction of FS in region $k\sim (\pi ,0)$
has been observed in ARPES data \cite{4,5}.
Large summary intensity of photoemission
from this region of $k$ have really been observed in \cite{12} for BSCCO.
However it occurs visible only at the photon energy
22 eV, but not at 33 eV.
 In our picture of split Hubbard bands the VHS of LHB is more strong
than that expected from tight-binding unsplit band. So additional
increase of intensity at $k\sim (\pi,0)$ is possible. But the
singlelayer models of cuprates cannot describe the observed dependence
of additional photoemission on the photon energy \cite{12}. As in
discussions about tunnel spectra, we come to necessity of extending the
study on bilayer models.

In conclusion, fine structure of upper edge of LHB in t-t'-U Hubbard
model and variational correlated state with VB formation can
describe some of low-energy phenomena in cuprates. Doping and temperature
dependencies of gap anisotropy obtained from photoemission data \cite{11}
are explained by interplay of gap and pseudogap
which is working function for electron removing from dielectric segments
of generalized FS. It is shown that only SC gap, but not pseudogap, is seen
in tunnel spectra. At $T\r0$  differential tunnel current for optimal doping
is similar to observed one, but it displays too sharp asymmetry
in heavily UD (and OD) regimes with peaks at negative (or positive)
voltages caused by normal state VHS. Model seems to need extension on
bilayer models.  Occupation number $n_k$ and spectral functions calculated
in MF approximation confirm the sharp Fermi boundary at $k_x=k_y$ direction
and smoothed (destructed ) Fermi boundary near $(\pi ,0)$ as it observed
in ARPES \cite{4,5}. Change of FS topology at optimal doping  and large
partly electron-like FS at OD case in our model seem to receive
confirmation in reexamined  ARPES data \cite{12}, but single-layer model
cannot explain the dependence of photoemission spectra on photon energy.

Work is supported by Russian Fundamental Research Foundation under Grants
No.7-03-33727€ and No.96-15-97492. One Authors is acknowledge MPI PCS for
kind hospitality. Authors are acknowledge stimulating
discussions with V.Ya.Krivnov.

%\newpage

\newpage{\bf Captions to Figures}
\vspace {0.15in}

Fig.~1.

a) Doping dependence of SC transition temperature $T_c (\d)$ in unit
[0.01t]. b)Double shift of photoemission edge $2\D_M(\d)$ at $k=(\pi,0)$
(curve 1) and corresponding SC gap $2W_k(\pi ,0)$ (curve 2) in same units.
c) The ratios $\xi=v_\D /\D_M$ (curve 1) and
$\xi_W=v_\D/W_k(\pi ,0)$ (curve 2) as function of doping.
d) Angular dependence of gap $\D(\f)$ for various temperatures
in heavily UD model $\d=0.14, ~~T_c/T_c^{max}=0.55$. Rest parameters
of model are in text.

Fig.~2. Differential tunnel current for nearly optimal doping $\d=0.2$
($T_c/t=0.0114$) at temperatures $T/t=0.002$, $0.014$, $0.018$.
Insets: the same and the current $I(V)$ for $T/t=0.002$, $0.018$ in
large voltage regions. Parameters of model are $U/t=8,~t'/t=0.05,~V/t=0.1$.

Fig.~3. Tunnel current $dI/dV$ at $T/t=0.002$, $0.014$, $0.018$  for
UD and OD models with $T_c/T_c^{max}=0.743$ and $0.63$ correspondingly.
Model parameters are same as in Fig.~2

Fig.~4. DOS of LHB for the MF AF state (curve 1) or for spiral states
with $Q_2=(0.9\pi ,\pi )$ and $Q_3=(0.93\pi ,0.93\pi )$ (curves 2 and 3).

Fig.~5. Occupation number $n_k$ (solid lines) and $A(k,\o =0,\g )$ with
$\g=0.003t$ (dotted lines) for MF AF state and spiral states  of UD and OD
models. Vector $k$ runs along path $Y-\G -M-Y$. For spiral states
vectors $Q$ are determined by values $q'=0.8\pi$ and $q=0.86$.
Dashed dotted curves on segments $\G -M-Y$ (upper grafics) correspond
to  AF state at nearly "optimal" doping $\d =0.3$.


\begin{thebibliography}{99}


\bibitem{1}
T.Timusk, B.Statt, Rep.Progr.Phys. {\bf 62}, 61 (1999)

\bibitem{2}
Z.M.Shen, D.S.Dessau, Phys.Rep., {\bf 253}, 1 (1995)

\bibitem{3}
R.S.Markievicz, J.Phys.Chem.Solids {\bf 58}, 1179 (1997).

\bibitem{4}
N.M.Norman, H.Ding, M.Randeria et al., Preprint cond-mat/97101163;

\bibitem{5}
N.M.Norman, Preprint cond-mat/9904048.

\bibitem{6}
T.Hasegawa, H.Ikuta, K.Kitazawa in Physical properties of High temperature
superconductors., ed.D.M.Hinzberg, v.III, p.525, 1992.

\bibitem{7}
Ch.Renner, B.Revaz, J.-Y.Genoud et al., Phys.Rev.Lett. {\bf 80}, 149 (1998).

\bibitem{8}
A.Matsuda, S.Sugita, T.Watanabe, Phys.Rev. {\bf B60}, 1377  (1999).

\bibitem{8a}
Y.DeWilde, N.Miyakawa, P.Guptasarma et al., Phys.Rev.Lett.{\bf 80},
153 (1998).

\bibitem{9}
J.M.Harris, P.J.White, Z.X.Shen et.al. Phys.Rev.Lett. {\bf 79}, 143 (1997).

\bibitem{10}
J.M.Harris, Z.X.Shen, P.J.White et.al., Phys.Rev. {\bf B 54}, R15665 (1997).
P.J.White, Z.-X. Shen, C.Kim et.al.,  Phys.Rev. {\bf B 54}, R15668 (1997).

\bibitem{11}
J.Mesot, N.R.Norman, MRanderia et al. Phys.Rev.Lett. {\bf 83}, 840 (1999).

\bibitem{12}
Y.-D.Chang, A.D.Gromko, D.S.Dessau et al., Phys.Rev.Lett. {\bf 83}, 3717
(1999).

\bibitem{13}
N.M.Plakida, V.S.Oudovenko, R.Horsch et al., Phys.Rev. {\bf B55}, (1997)
11997.

\bibitem{14}
A.A.Ovchinnikov, M.Ya.Ovchinnikova, Phys.Lett. {\bf A249}, 531 (1998).


\bibitem{15}
A.A.Ovchinnikov, M.Ya.Ovchinnikova, E.A.Plekhanov, Pisma JETP {\bf 67},
350 (1998); JETP {\bf 87},534 (1998); {\bf 88}, 356 (1999)
[Z.Eksp.Teor.Fiz.{\bf 114}, 985 (1998); {\bf 115}, 649 (1999)].

\bibitem{16}
R.O.Kuzin, R.Hayn, A.F.Barabanov et.al., Phys.Rev. {\bf B 58}, 6194 (1998).

\bibitem{17}
F.Onufrieva, P.Pfeuty and M.Kisilev, Phys.Rev.Lett {\bf B 82}, 2370 (1999).

\bibitem{18}
P.J.White, P.J.Shen, Z.X.Kim et al. Phys.Rev.{\bf B54} R15669(?) (1996).


\bibitem{19}
J.Schmalian, D.Pines, B.Stojkovic, Phys.Rev. {\bf B 60}, 667 (1999).

\bibitem{20}
T.Schneider, H.DeRaedt, M.Frick, Z.Phys.B: Cond.Matt. {\bf 76}, 3 (1989).

\bibitem{21}
Z.-X.Shen, J.R.Schriefer, Phys.Rev.Lett. (\bf{78}, 1771 (1997).


\end{thebibliography}
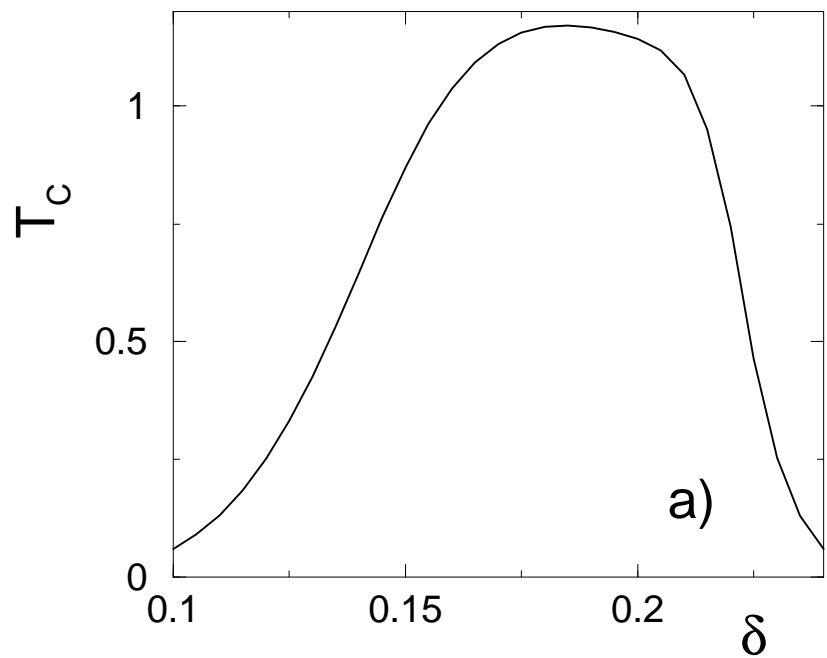
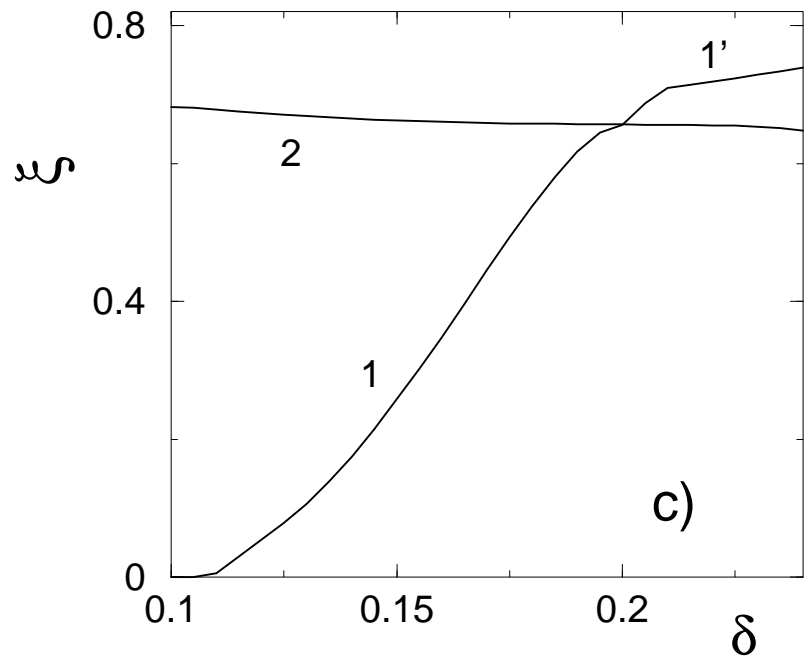
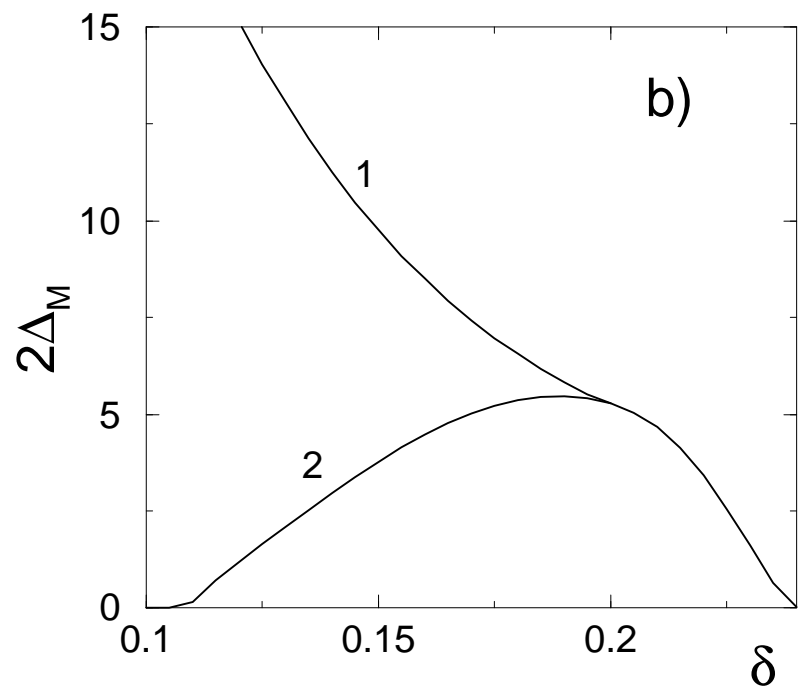
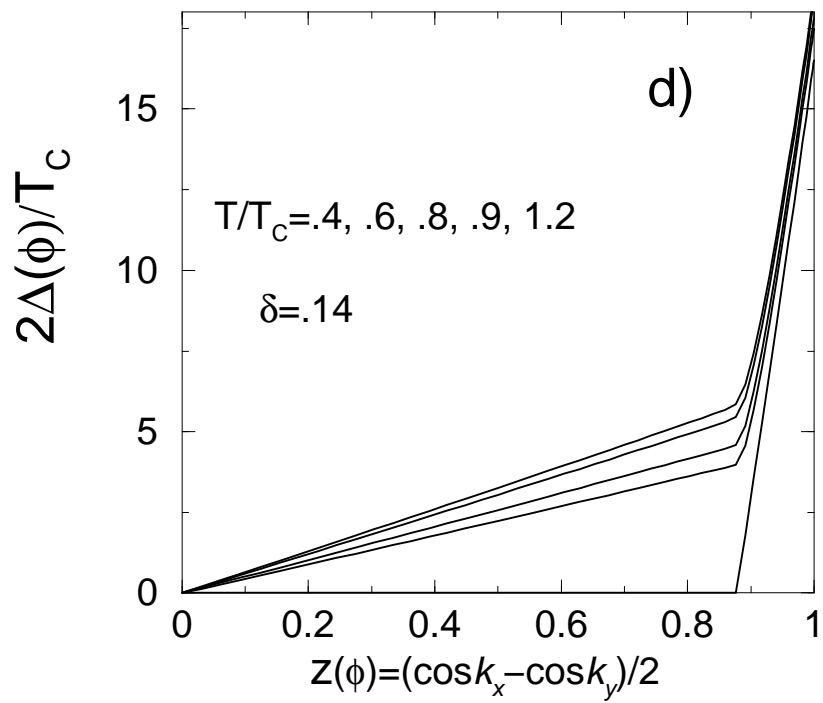
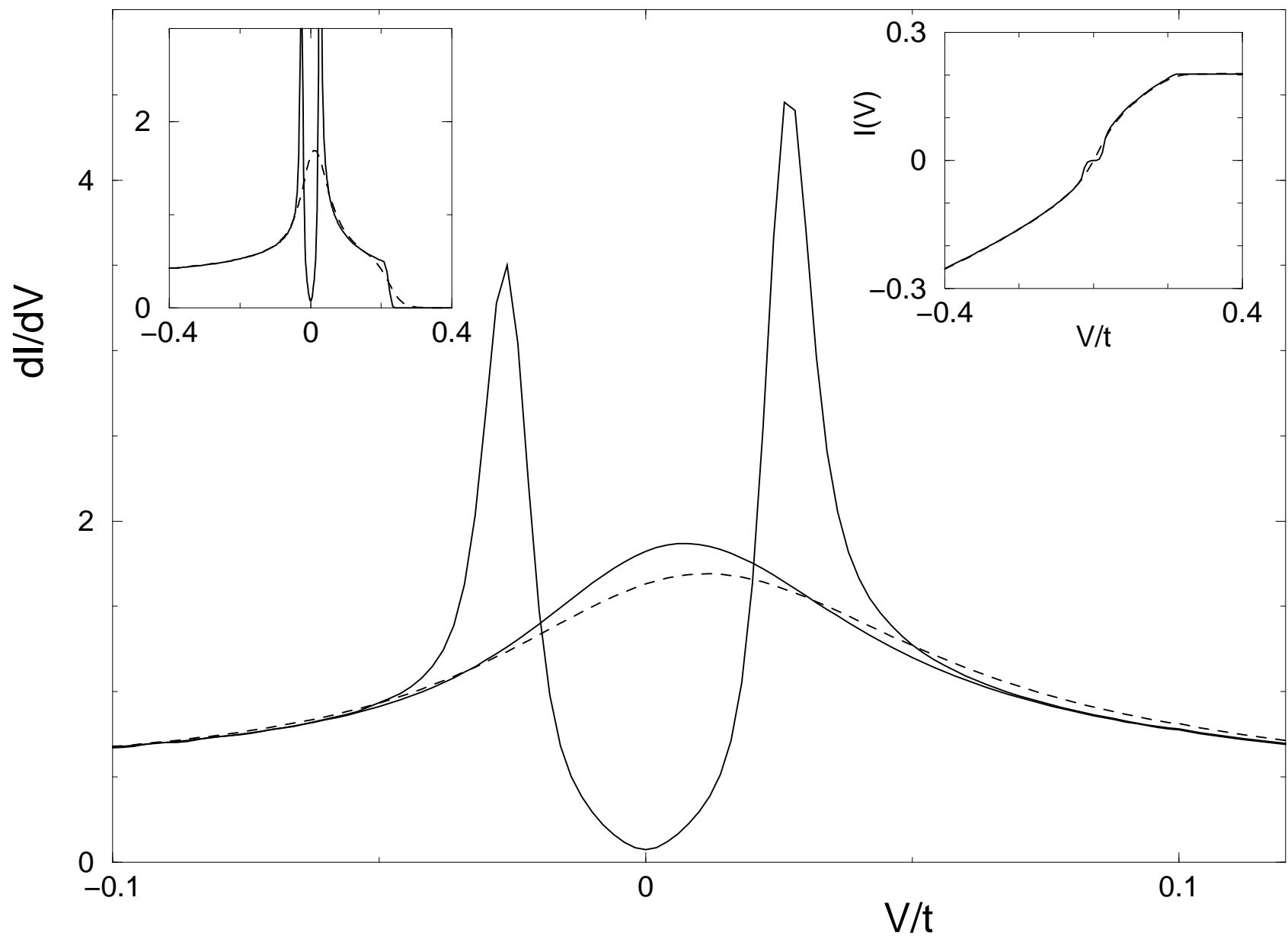
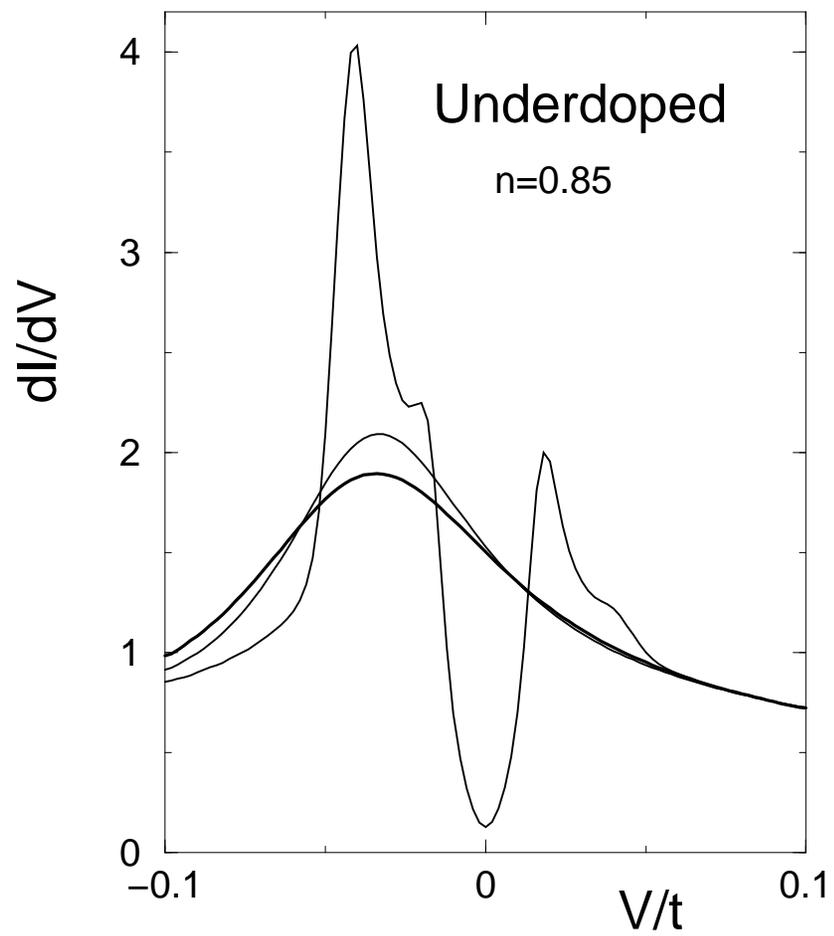
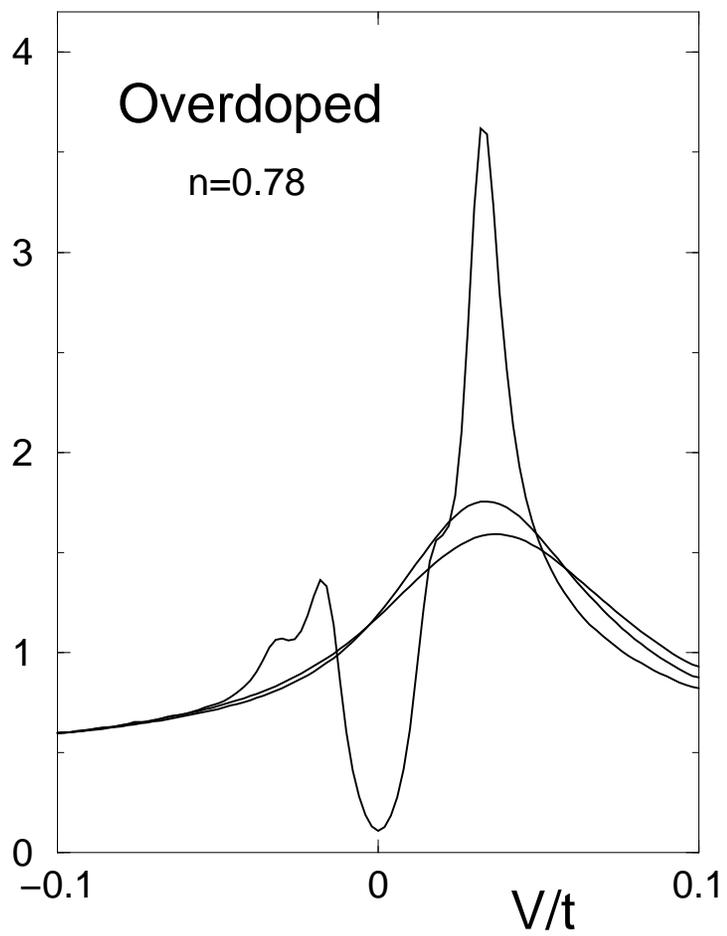
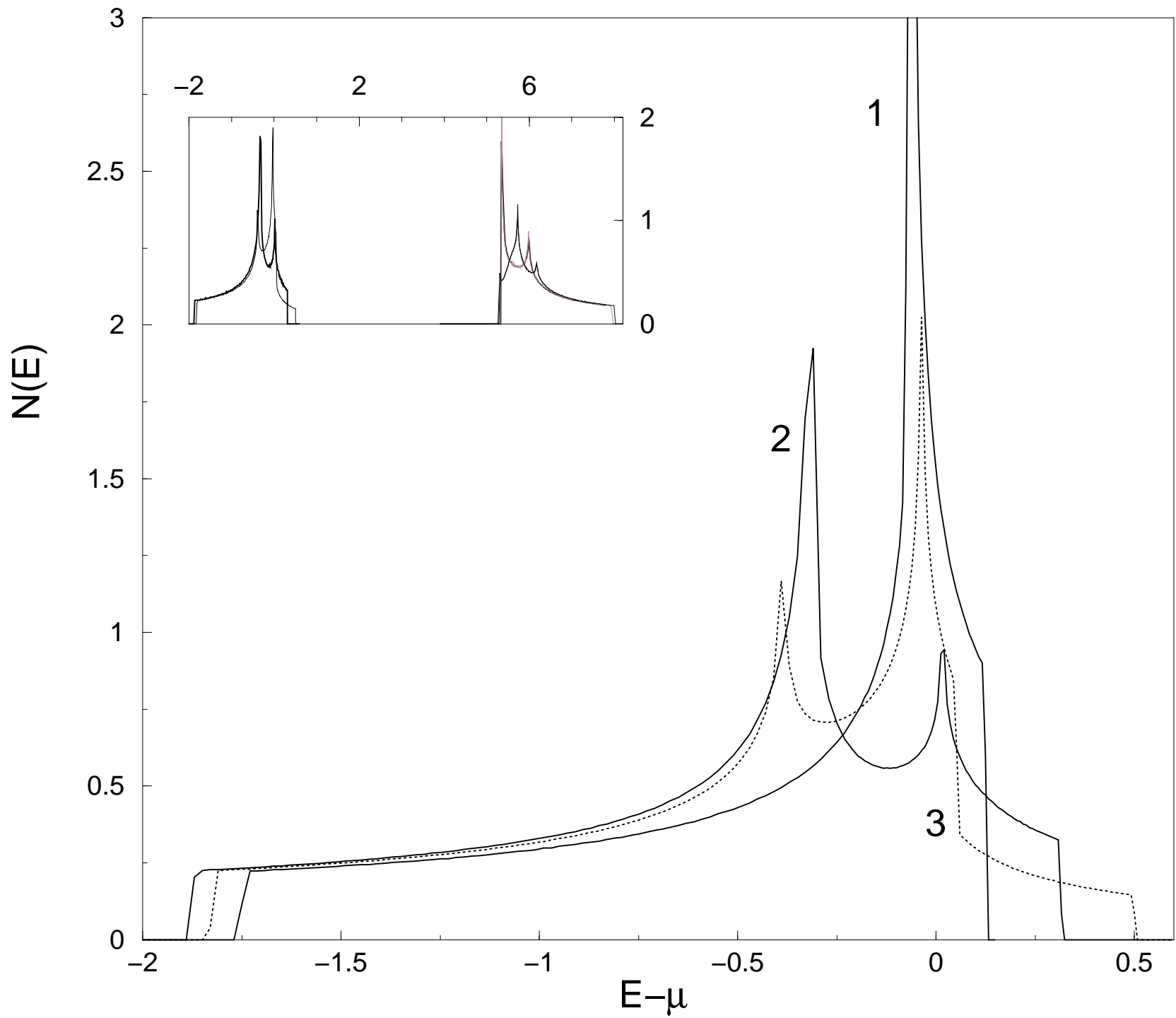
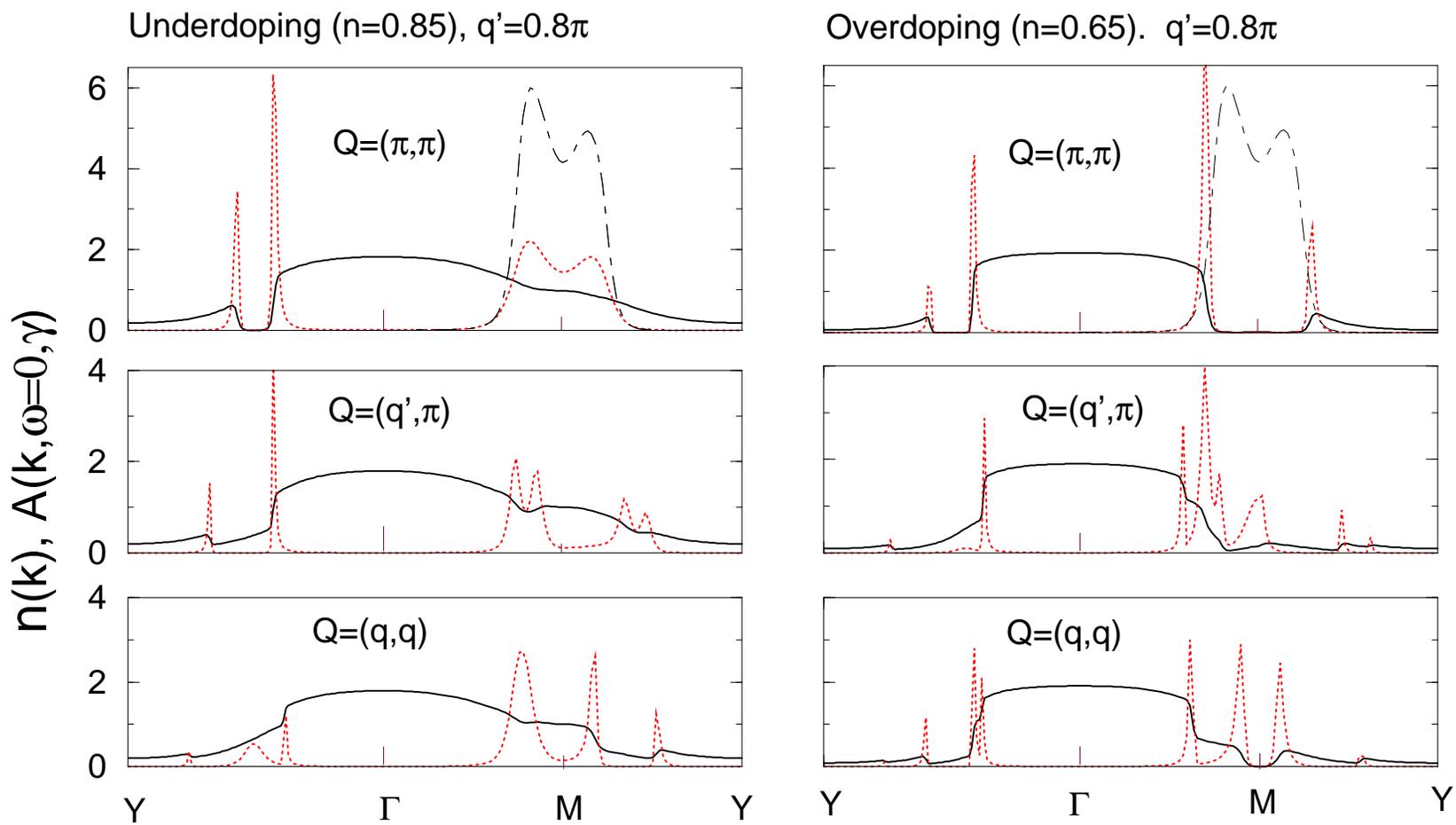
\end{document}